\documentclass[draft]{agujournal2019}
\usepackage{url} 
\usepackage{natbib}

\usepackage[inline]{trackchanges} 
\usepackage{soul}
\usepackage{amsmath}
\usepackage{setspace}
\renewcommand{\APACrefnote}[1]{}

\draftfalse

\journalname{Earth and Space Science}

\begin{document}
%
%

\title{Determining Undersampled Coastal Tidal Harmonics using Regularized Least Squares}

%
%




\authors{Ruo-Qian Wang*, Behzad Golparvar, Morgan Mark}


\affiliation{}{Department of Civil and Environmental Engineering, Rutgers, the State University of New Jersey, Piscataway, NJ, 08854}




\correspondingauthor{Ruo-Qian Wang}{email address:rq.wang@rutgers.edu}




\begin{keypoints}
\item A new algorithm is developed to determine tidal amplitudes with undersampled elevation data
\item Synthetic data experiment and a real case study validate the new algorithm
\item The new method to determine tidal amplitudes achieves a high accuracy
\end{keypoints}

%
%

%
%


\begin{abstract}
Satellite altimetry, which measures water level with global coverage and high resolution, provides an unprecedented opportunity for a wide and refined understanding of the changing tides in the coastal area, but the sampling frequency is too low to satisfy the Nyquist frequency requirement and too few data points per year are available to recognize a sufficient number of tidal constituents to capture the trend of tidal changes on a yearly basis. To address these issues, a novel Regularized Least-Square approach is developed to relax the limitation to the range of satellite operating conditions. In this method, the prior information of the regional tidal amplitudes is used to support a least square analysis to obtain the amplitudes and phases of the tidal constituents for water elevation time series of different lengths and time intervals. Synthetic data experiments  performed in Delaware Bay and Galveston Bay showed that the proposed method can determine the tidal amplitudes with high accuracy and the sampling interval can be extended to the application level of major altimetry satellites. The proposed algorithm was further validated using the data of the altimetry mission, Jason-3, to show its applicability to irregular and noisy data. The new method could help identify the changing tides with sea-level rise and anthropogenic activities in coastal areas, informing coastal flooding risk assessment and ecosystem health analysis.
\end{abstract}

\section*{Plain Language Summary}
Using satellite data to directly measure the difference from high tide to low tide is a long-lasting challenge because the interval between satellite visits to the same location is too long and there are too few data points per year that can be used. We developed a new algorithm to enable the direct measurement of coastal tidal activities. The method is validated using observations. The tests showed that the new algorithm outperforms its peers and meets the application requirements. The new algorithm paves a way to understand the global changes in tidal motion, which could help governments and coastal communities analyze the changing risk of coastal flooding and ecosystems and support informed decisions to prepare for and mitigate the impacts.

%
%

%


%
%
%
%

\section{Introduction}
Coastal tides have been experiencing ubiquitous changes in the past century due to anthropogenic activities (e.g., dredging and land reclamation) and environmental processes (e.g., erosion and sea-level rise) \citep{haigh_tides_2020, talke_changing_2020}. These changes may potentially increase the risk of property damage, loss of life, ecosystem degradation, and environmental injustice. A remarkable example is the increase of high tide flooding (also called “blue sky” flooding or “sunny day” flooding), which is shallow but widely spread flooding which has water levels ranging from about 0.5 meters (m) to 0.65 m above the mean higher high water (MHHW) level \citep{sweet_patterns_2018}. NOAA (National Ocean and Atmospheric Agency) reported that 75\% of the US East and Gulf Coast’s monitored locations witnessed an increasing trend of high tide flooding \citep{sweet_2019_2020}, which disrupts transportation, sewage, and other infrastructure systems, devalues real estate, reduces income and jobs, exposes health hazards, increases public health risk, increases groundwater salinity, and deteriorates coastal ecosystems \citep{moftakhari_cumulative_2017}. These relatively more frequent, smaller floods, at some locations, may be proved more costly than large, infrequent extreme events \citep{moftakhari_cumulative_2017, li_evolving_2021}.

Coastal tidal changes have been understudied until the recent observation of rapidly evolving tidal amplitudes in some estuaries (see reviews in \citet{haigh_tides_2020} and \citet{talke_changing_2020}). For example, the tidal range at Albany, New York, has been increasing by 0.5 cm per year since 1920 but has become steady in recent years. Similar observations have been made at Philadelphia, Wilmington, and Jacksonville. The secular (i.e., nonperiodic and long-term) tidal range changes are accompanied by the changing tidal phases, e.g., a decrease of about 30 degrees was found in Philadelphia \citep{ross_fingerprints_2017}. But tidal ranges remained the same level in other estuaries after removing the 18.61-year nodal cycle, e.g., at Sandy Hook in New Jersey and Boston in Massachusetts. Changing tides can alter fundamental coastal processes, e.g., the turbulent mixing changed between 1954 and 2005 in the Ems estuary in the Netherlands \citep{de_jonge_influence_2014}, and the overtide (M4) decreased with the changing tides in Argentina \citep{santamaria-aguilar_sea-level_2017}. To date, the causes that lead to the varying rates from place to place are unclear, and there has been an increasing interest in the changes in tides and the observational evidence in support of the observation.


High-quality and widely-coverage data is demanded to fully recognize and systematically examine the change of coastal tidal dynamics, but data scarcity remains an ongoing issue for coastal monitoring programs \citep{barnard_coastal_2015, pianca_shoreline_2015, turner_multi-decade_2016}. Most of these studies rely on data from tide gauges \citep{parker_tidal_2007}, which are sparsely distributed across the globe and make it impossible to obtain large-scale patterns \citep{devlin_extended_2019}. 

Satellite altimetry technology, which sends radar pulses and receive the echoes reflected from the Earth's surface to measure sea surface heights, has now reached the accuracy of centimeters within 5 km from the shoreline \citep{cipollini_monitoring_2017,valle-rodriguez_poleward_2017, passaro_cross-calibrating_2016,birgiel_examining_2018}. Multiple aspects of the technological advance drive the technological improvement such as new platforms with proven altimetry precision and accuracy and new and dedicated data processing schemes \citep{passaro_ales_2014,passaro_ales_2018}. The breakthrough of coastal altimetry technology empowered a series of applications, e.g. the estimation of ocean currents \citep{jebri_exploiting_2016,jebri_interannual_2017,salazar-ceciliano_coastal_2018,valle-rodriguez_poleward_2017}, inland flow \citep{gleason_remote_2020,biancamaria_satellite_2017}, tidal mixing fronts \citep{dong_identification_2018}, tidal energy dissipation \citep{egbert_estimates_2001}, storm surge heights \citep{ji_observing_2019}, and vertical land motion \citep{oelsmann_zone_2020}. Because this technology is free of clouds or significant weather impacts \citep{wang_recommender_2022}, it is considered as an ideal tool to continuously monitor the Earth's surface. 

Satellite altimetry has been used in estimating ocean tides. For example, \citet{fang_empirical_2004} studied the tidal amplitudes in East Asia and \citet{ray_measurements_2016} analyzed the tides in Hudson Bay, Canada. Due to the poor temporal resolution of altimetry satellites, they must use ten years of data at the cross-over points of two different scanning passes to achieve acceptable results. In addition, the analysis was limited to offshore regions and could not capture tidal changes. \citet{bij_de_vaate_secular_2022} is probably the only study focusing on tidal changes over different years. Although their method relaxed the data length requirement to 4 years and allowed the analysis of altimetry data other than cross-over points, the application is limited to the deep ocean and the determined tidal amplitudes were not  validated despite the changing trend being compared with gauge data.

All these studies were based on Harmonic Analysis (HA) to determine tidal amplitudes, but this method is known to be inaccurate in analyzing measurement data of low temporal resolution, especially in the underdetermined situation where observation data points are less than unknowns to determine. Various new methods were developed with different assumptions of the residual of harmonic analysis and compared in \citet{innocenti_analytical_2022}. In addition, machine learning methods are applied to improve the performance of HA \citep{gan_application_2021,su_prediction_2023}. Not surprisingly, they showed improved performance in reproducing tidal signals attributed to a large number of tuning parameters in machine learning models, but they could not convincingly extract the amplitudes of different tidal constituents to inform the physics of tidal processes. More importantly, no method mentioned above is shown feasible to be extended to extract tidal amplitudes from the observation data at the temporal resolution of altimetry satellites, e.g., 9.9 days of Jason-3.

\begin{figure}
    \centering
    \includegraphics[width=\linewidth]{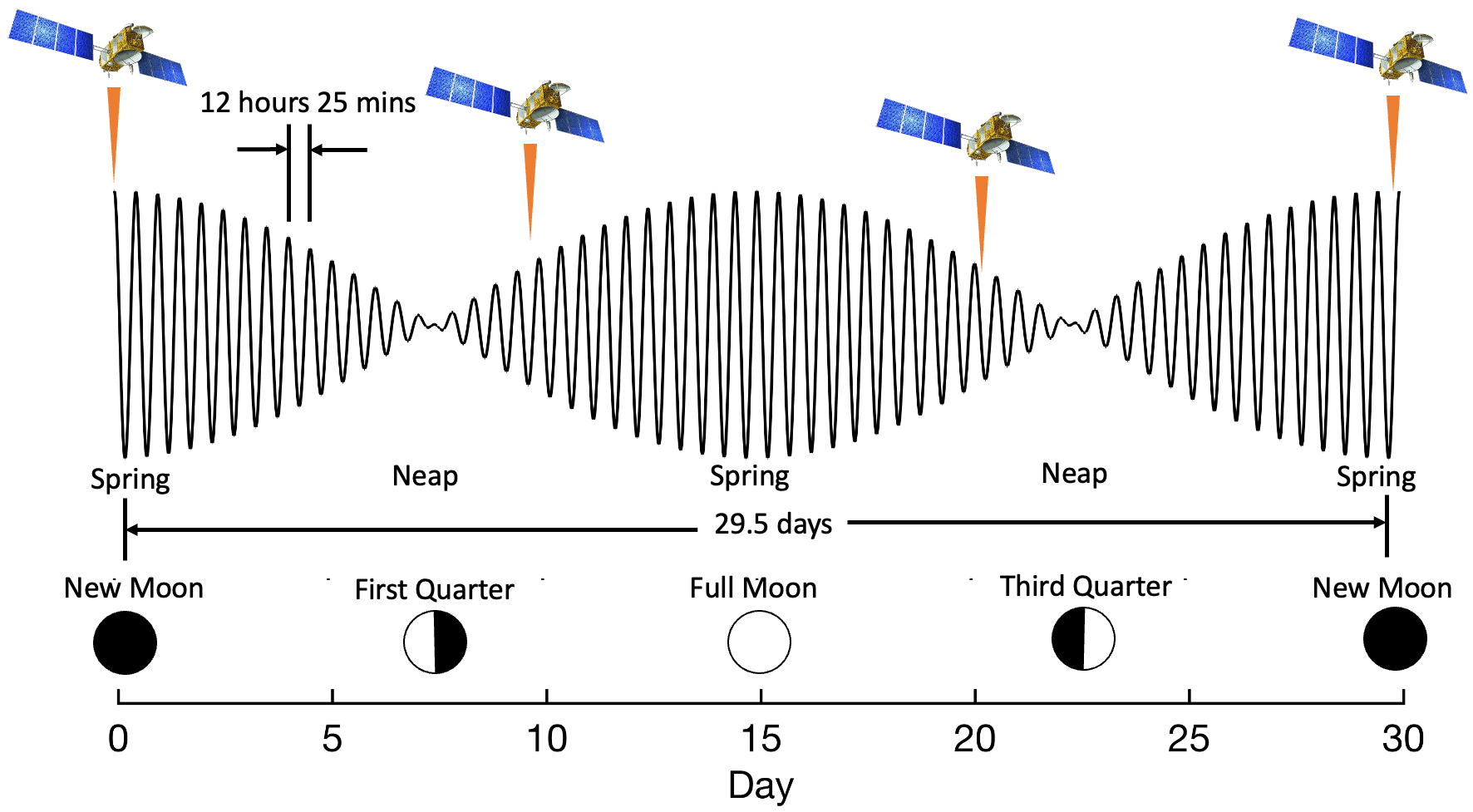}
    \caption{Altimetry satellite Jason-3 measures water elevation every 9.9 days, which cannot resolve the tidal amplitudes with a typical period of 12 hours and 25 mins according to the Nyquist principle.}
    \label{fig:spring}
\end{figure}

The major bottleneck preventing the tidal amplitude analysis is the difficulty of extracting tidal amplitudes with the temporal resolution of altimetry satellites -- the long revisit time cannot support the data requirement of HA (Figure \ref{fig:spring}). Specifically, Sea Surface Height (SSH), $h_H$, is assumed to follow a series of harmonic series in HA, i.e.
\begin{equation}
    h_H=m+a t+\sum_{k=1}^n[A_k f_k cos(\omega_k t+\phi_k+u_k )] ,	
\end{equation}
where $t$ is time, $a$ is the slope of the mean sea level change, $m$, $\omega$, $\phi$, $f$, and $u$ are the mean water level, the frequency, the astronomical phase angle, and the nodal corrections for the amplitude and phase of each tidal constituent, and $n$ is the number of tidal constituents \citep{le_provost_chapter_2001}. Popular tools such as UTide \citep{codiga_unified_2011} and Pytides \citep{cox_pytides_2022} are available in Matlab and Python to perform the harmonic analysis  to determine the amplitudes and phases of tidal constituents from sea level data collected at irregular times using a Lomb-Scargle least-square spectral approach \citep{press_fast_1989}.

This traditional HA method is limited by the data requirements of the Rayleigh criterion and the Nyquist–Shannon sampling theorem. The Rayleigh criterion requires a sufficiently long period of data to differentiate tidal constituents of close frequencies and it was shown that data of one year is necessary to recognize 39 key tidal constituents \citep{parker_tidal_2007, godin_resolution_1970}. The Nyquist–Shannon sampling theorem dictates that the sampling frequency must be greater than two times the frequency of the targeted phenomenon. In tidal analyses, this means that the sampling interval must be shorter than 6.2 hours to capture M2, the dominant constituent in most places of the world, in spite that least-square based HA can, to a limited range, override the requirement in practice. Unfortunately, the frequency of altimetry satellites is much lower than required in practice and tidal amplitudes cannot be accurately estimated, e.g., the major altimetry satellite of Jason-3 revisits the same location of the earth's surface every 9.9 days (Figure \ref{fig:spring}).

Recognizing the gap between the satellite measurement and harmonic analysis, scientists developed a data-science method to resolve the issues: An interpolation method called Constrained Harmonic Analysis (CHA) was proposed to estimate tidal amplitudes, where data is fitted into two harmonic series of neighboring tidal gauges \citep{matte_constrained_2018}. By assigning a series of weights between 0 to 1 (“0” means the measured tidal harmonics is the same as the first gauge and ``1'' means the second), tidal amplitudes and phases can be ``interpolated'' between two reference gauges to fit the observation. A good agreement with tidal gauge data was shown \citep{matte_constrained_2018}. However, this method cannot reach the necessary frequency to enable the satellite-based assessments of tidal amplitudes (see details in Section \ref{sec:results}).

Consequently, the conventional HA requires processing segments of several years of data, while to study secular changes in tides, we would want to retrieve tidal constants on a yearly basis (as is done in most tidal change studies using tide gauge data). For that reason, the authors propose a novel approach that allows the processing of shorter records of satellite data with poor temporal resolution. In Section \ref{sec:method}, we describe a novel regularized least-square approach and explain the rationale behind it. Section \ref{sec:results} showed the results of synthetic data experiments to validate the method and a demonstration of its application in processing satellite altimetry data.

\section{Methods}\label{sec:method}

\subsection{Regularized Least-Square Harmonic Analysis}
We propose an algorithm called Regularized Least-Square Harmonic Analysis (ReLSHA) to address the strict data requirements posed by the Rayleigh criterion and Nyquist–Shannon sampling theorem. Regularized Least-Square has been used in compressed sensing to reconstruct a signal compressible by a known transform (e.g., wavelet) with prior information. This method allows the measurements of undersampled signals \citep{eldar_compressed_2012,tsaig_extensions_2006}. Numerous examples demonstrated that such compressed sensing methods can relax the limit of Nyquist–Shannon sampling limit \citep{eldar_compressed_2012,tsaig_extensions_2006,mclean_nyquistovercoming_2005}. Specifically, the algorithm is designed to minimize an objective function,
\begin{equation}
\min_{A_k f_k,\phi_k+u_k} J=(1-\lambda) \left \| h_H (A_k f_k,\phi_k+u_k) -h_0  \right \|+\lambda \left \| A_k^2 f_k^2-A_{0,k}^2\right \|,
\end{equation}\label{eq:relsha}
where $h_H$ is the time series reconstructed by the harmonics, $h_0$ is the time series of the measured water level, $\lambda$ is the regularization weight, $A_{k}$ are the tidal amplitudes of the harmonics, $A_{0,k}$ are the amplitudes of a reference station (a good choice of the reference site is the nearest tidal gauge), $f_k$ are the node factors of the lunar nodal cycle, $\phi_k$ are the major phases of the tidal constituents, $u_k$ are slowly changing angles depending upon the 18.6-year lunar nodal cycle, $k=1,2,\cdots n$, and $n$ is the targeted number of constituents. In this study, we assign $n=37$ to be consistent with the number of the major constituents used in NOAA tidal gauges. The strategy of this algorithm is thus to determine the amplitudes and phases of the harmonic series to ``best fit'' the measured water level with a penalization to the amplitude difference between the determined harmonics and the reference. As a main effect, regularized least-square reduces the impact of local minimums in the optimization process and improved the convergence to the global optimal. 

Note that only the amplitude difference of the harmonic series is penalized due to several benefits in the numerical scheme and its application: because the tidal phases are involved in the nonlinear trigonometrical functions, determining tidal phases is sensitive to the setup of the numerical scheme including the starting point, noise of the data, and the method of optimization. A potential solution to this problem is to regularize tidal phases in a similar approach as the tidal amplitudes. But such regularization will lead to a nonlinear objective function and deriving the Jacobians will be almost impossible, which would result in slow computation and compromise the possibility and effectiveness of convergence. In addition, we found the reference tidal phase data collected from the internet and other data sources overseas often depends on time zones and contain incorrect or lack metadata, so including inaccurate or lacking phases will prevent the application of the new method if tidal phases are included in the regularization. Therefore, we decide not to analyze or validate tidal phases in the new method, although they are determined through this algorithm.

\subsection{Computational Scheme}
Since the harmonic series involves nonlinear trigonometric functions to design efficient solution procedures, we developed a new Regularized Least-Square strategy to linearize the minimization problem to accelerate the computation and improve the solution convergence. First, the mean and linear components of the measured water level are removed. Then, Equation \ref{eq:relsha} is linearized as following

\begin{align}
\begin{split}
    J &=(1-\lambda) \left \| \sum_{k=1}^n [A_k f_k cos(\omega_k t_i+\phi_k+u_k )] -h_{0,i}  \right \|+\lambda \left \| A_k^2 f_k^2-A_{0,k}^2\right \| \\
    &=(1-\lambda) \left \|\sum_{k=1}^n [ A_k f_k cos(\phi_k+u_k)cos(\omega_k t_i) - A_k f_k sin(\phi_k+u_k)sin(\omega_k t_i)] -h_{0,i}  \right \| \\
    &\quad +\lambda \left \| A_k^2 f_k^2-A_{0,k}^2\right \|\\
    &=(1-\lambda)  (\mathbf{H}\mathbf{x} - \mathbf{h})^T (\mathbf{H}\mathbf{x} - \mathbf{h})+\lambda (\mathbf{K}(\mathbf{x}\bigodot \mathbf{x})- \mathbf{q})^T [\mathbf{K}(\mathbf{x}\bigodot \mathbf{x})- \mathbf{q}],\\
\end{split}
\end{align}
where $\bigodot$ is the Hadamard product sign, which performs the elementwise multiplication, the harmonic matrix $\mathbf{H}$ is
\begin{align}
    \mathbf{H}=\begin{bmatrix}
 &cos(\omega_1 t_1)  &cos(\omega_2 t_1)  & \cdots & cos(\omega_n t_1)  &sin(\omega_1 t_1)  &sin(\omega_2 t_1)   & \cdots & sin(\omega_n t_1)\\ 
 &cos(\omega_1 t_2)  &cos(\omega_2 t_2)  & \cdots & cos(\omega_n t_2)  &sin(\omega_1 t_2)  &sin(\omega_2 t_2)   & \cdots & sin(\omega_n t_2)\\ 
 & \vdots  & \vdots  & \ddots  &\vdots  & \vdots & \vdots & \ddots  &\vdots \\ 
&cos(\omega_1 t_m)  &cos(\omega_2 t_m)  & \cdots & cos(\omega_n t_m)  &sin(\omega_1 t_m)  &sin(\omega_2 t_m)   & \cdots & sin(\omega_n t_m)\\
\end{bmatrix},
\end{align}
the vector $\mathbf{x}$ to solve is
\begin{align}
\mathbf{x}=\begin{bmatrix}
&A_1 f_1 cos(\phi_1+u_1)\\ 
&A_2 f_2 cos(\phi_2+u_2)\\ 
&\vdots\\ 
&A_n f_n cos(\phi_n+u_n)\\ 
&A_1 f_1 sin(\phi_1+u_1)\\ 
&A_2 f_2 sin(\phi_2+u_2)\\ 
&\vdots\\ 
&A_n f_n sin(\phi_n+u_n)\\
\end{bmatrix},
\end{align}
the measured water level vector, $\mathbf{h}$, is
\begin{align}
\mathbf{h}=\begin{bmatrix}
&h_{0,1}\\ 
&h_{0,2}\\ 
&\vdots\\ 
&h_{0,m}\\ 
\end{bmatrix},
\end{align}
the transformation matrix, $\mathbf{K}$, is used to extract tidal amplitudes, i.e.,
\begin{align}
    \mathbf{K}=\begin{bmatrix}
  1 &  &  &  & 1 &  & &\\ 
   & 1 &  &  &  &  1& &\\ 
   &  & \ddots &  &  &  &\ddots &\\ 
   &  &  & 1 &  &  & &1
\end{bmatrix},
\end{align}
and the reference vector, $\mathbf{q}$ is
\begin{align}
\mathbf{q}=\begin{bmatrix}
&A_{0,1}^2\\ 
&A_{0,2}^2\\ 
&\vdots\\ 
&A_{0,n}^2\\ 
\end{bmatrix}.
\end{align}

This linearized function is convenient to solve. Once $\mathbf{x}$ is obtained, the tidal amplitudes $A_k$ can be generated using
\begin{equation}
    A_k=\sqrt{(\mathbf{K}(\mathbf{x}\bigodot \mathbf{x}))_k}.
\end{equation}

To promote the quick convergence, the Jacobian of $J$ was derived, i.e.
\begin{equation}
    \frac{\mathrm{d} J}{\mathrm{d} \mathbf{x}}=2(1-\lambda)\mathbf{H}^T(\mathbf{H}\mathbf{x} - \mathbf{h})+4\lambda{Diag}(\mathbf{x})[\mathbf{K}(\mathbf{x}\bigodot \mathbf{x})- \mathbf{q}],
\end{equation}
where $Diag(\mathbf{x})$ is a function to convert the vector $\mathbf{x}$ into a diagonal matrix.

\subsection{Testing Data}
\begin{figure}
    \centering
    \includegraphics[width=\linewidth]{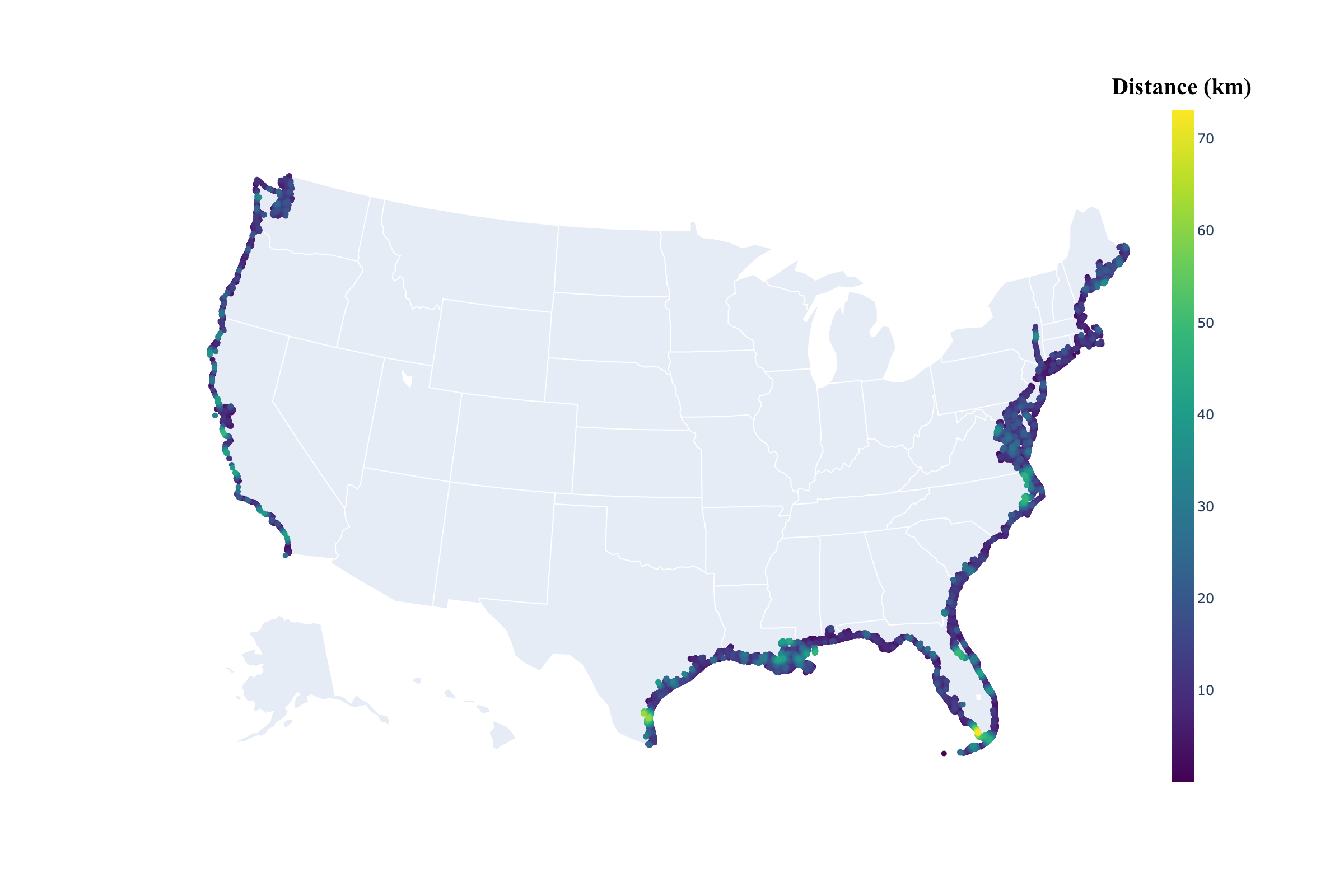}
    \caption{A survey of the distance to the nearest NOAA tidal harmonic reference site for any point along the Contiguous United States shoreline.}
    \label{fig:distance}
\end{figure}

Since this proposed ReLSHA method requires a reference vector of tidal amplitudes, a rational choice would be the data from the nearest tidal gauge. 
We performed a geographic study and found for any point along the coastline of the Contiguous United States one can always find a reference tidal gauge within $\sim$73 km (Figure \ref{fig:distance}). Note that the longest distance was found in South Florida and a few places in southern California are also close to the maximum distance.

\begin{figure}
    \centering \includegraphics[width=\linewidth]{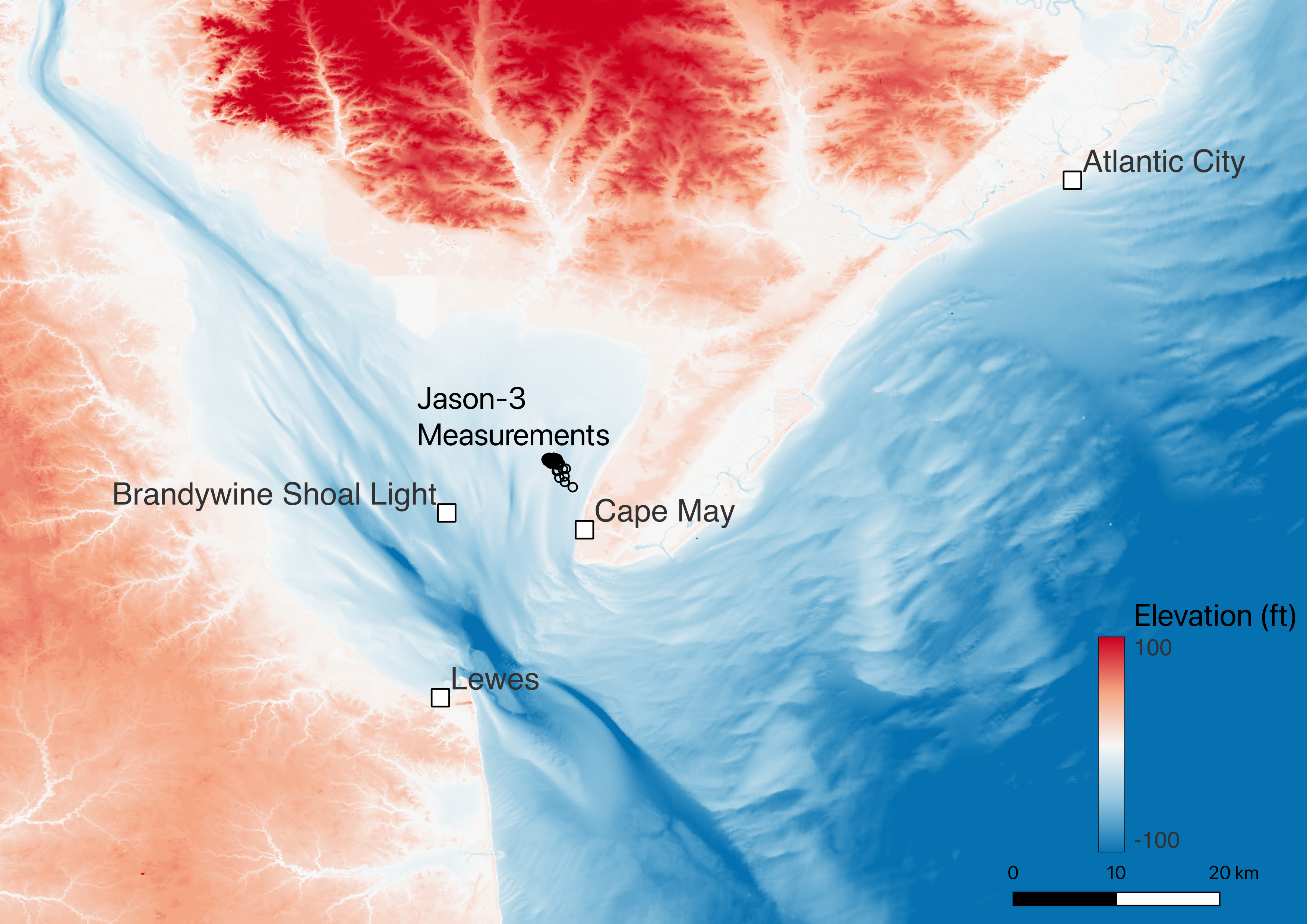}
    \caption{The map of the NOAA tidal gauges and the location of the Jason-3 measurements near Cape May used in this study.}
    \label{fig:sites}
\end{figure}

The first synthetic data experiment was conducted to validate the method and reveal the tidal amplitude changes in Delaware Bay and the performance of the new scheme is compared with the original HA and CHA. We calculated the 37 major tidal constituents using the observational data of tidal levels in 2021 from the NOAA station, Brandywine Shoal Light (BSL). Note that the testing data includes the non-tidal signal such as storms. The tidal amplitudes at Atlantic City (AC) were used as the reference. This reference site was deliberately chosen to set the distance at the level of 73 km and the two sites are located in different hydrodynamic environments -- BSL is inside Delaware Bay and AC faces the open ocean (Figure \ref{fig:sites}).

The tidal amplitudes and phases of two neighboring NOAA sites at Cape May, NJ and Lewes, DE were used as the reference sites for the input of CHA. The NOAA’s time series of the water level data collected every 6 mins was resampled with a range of intervals from 12 mins to 11 days and the total length was cut randomly to prepare the data of various sampling intervals and total lengths to explore how much the Nyquist-Shannon theorem and the Rayleigh criterion can be relaxed. The tidal amplitudes of the 37 constituents at BSL provided by NOAA were used as the ground truth (denoted as $A_{k,true}$) to evaluate the performance of the methods – the Relative Root Mean Square Error (RRMSE) of the 37 tidal constituents, $RRMSE=\frac{\sqrt{\frac{1}{n}\sum_{k=1}^n(A_k-A_{k,true} )^2}}{\sum_{k=1}^n A_k}\times100\%$, is used to compare the performance among different algorithms.

\begin{figure}
    \centering \includegraphics[width=\linewidth]{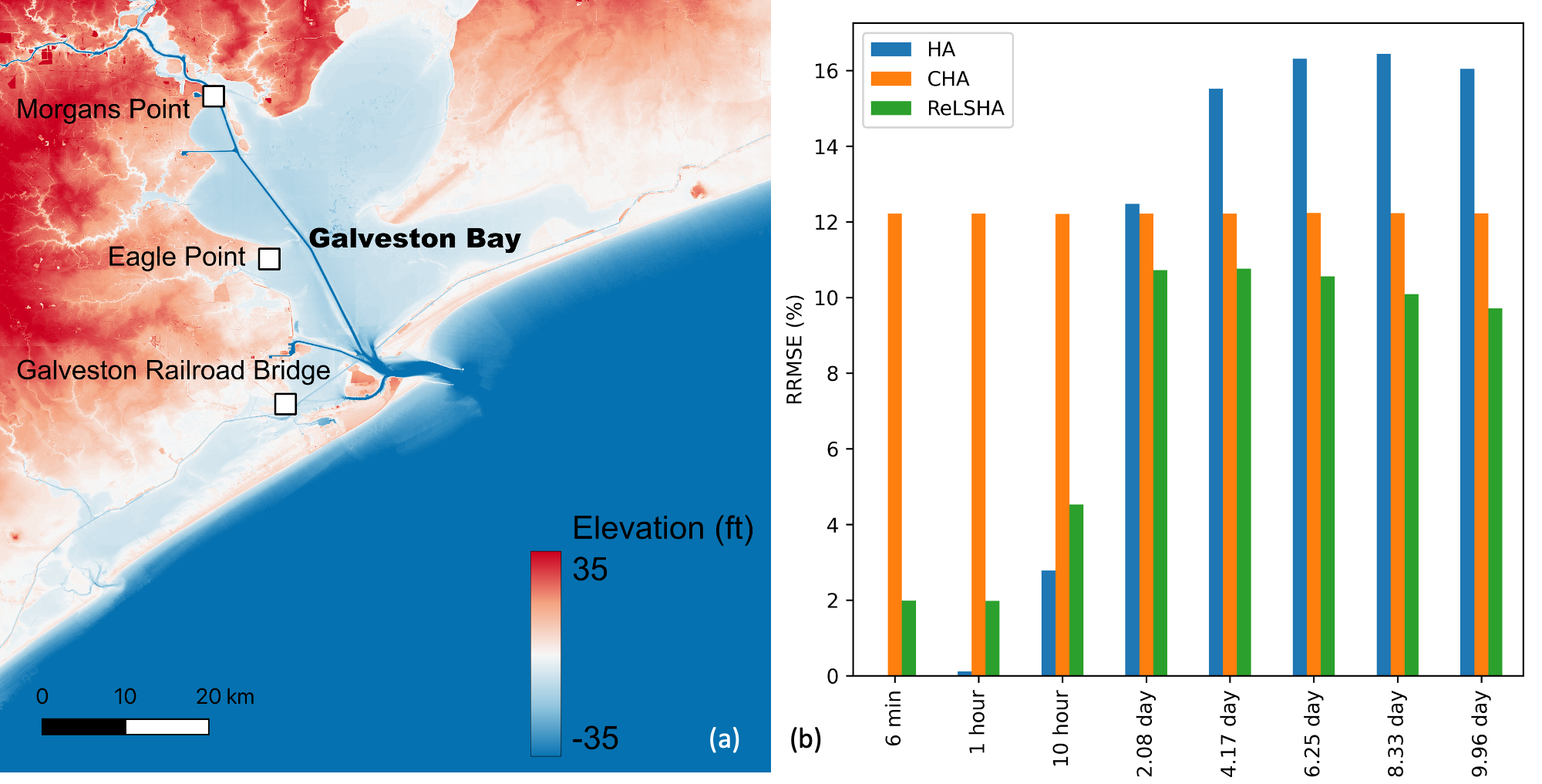}
    \caption{(a) The map of the NOAA tidal gauges used in the synthetic data experiment in Galveston Bay, TX; (b) the error of different methods in determining the tidal amplitudes for different sampling intervals.}
    \label{fig:galveston}
\end{figure}

The second synthetic data experiment was conducted in Galveston Bay, Texas (Figure \ref{fig:galveston}). Again, 37 major tidal constituents were calculated for Eagle Point using the observational data of tidal levels for 20 years from 2002 to 2021. Since diurnal tides dominate the tidal dynamics in Galveston Bay, the second site was selected to differentiate from the first to further test the validity of the algorithm. For ReLSHA testing, the tidal amplitudes at Galveston Railroad Bridge were used as the reference. For CHA, tidal data from the two neighboring sites, i.e., Galveston Railroad Bridge and Morgans Point, were used to support the analysis. The ground truth data for comparison was obtained by performing the harmonic analysis of the Eagle Point gauge data with a 6-min interval for every year from 2002 to 2021.

After the validation, ReLSH was applied to analyze the sea surface height (SSH) using the data of Jason-3. The data of the 1-Hz alongtrack altimeter data of Jason-3 was collected from Sea Level Anomalies Along-track Level-2+ Product (L2P) provided by the platform of Cnes AVISO-SALP (https://www.aviso.altimetry.fr/). The altimetry range data were extracted and corrected by dry and wet troposphere correction, and sea state bias correction (Adaptive Retracking) following the X-Track product manual. The measurements of Pass 288 close to the NOAA tidal gauge of Cape May were selected for analysis (Fig. \ref{fig:sites}). Note that some cycles have no data of Pass 288 due to data quality control or other reasons. In this validation task, BSL was used as the reference site and the tidal amplitudes based on NOAA tidal gauge at Cape May was considered as the ground truth.

\section{Results}\label{sec:results}
\begin{figure}
    \centering
    \includegraphics[width=\linewidth]{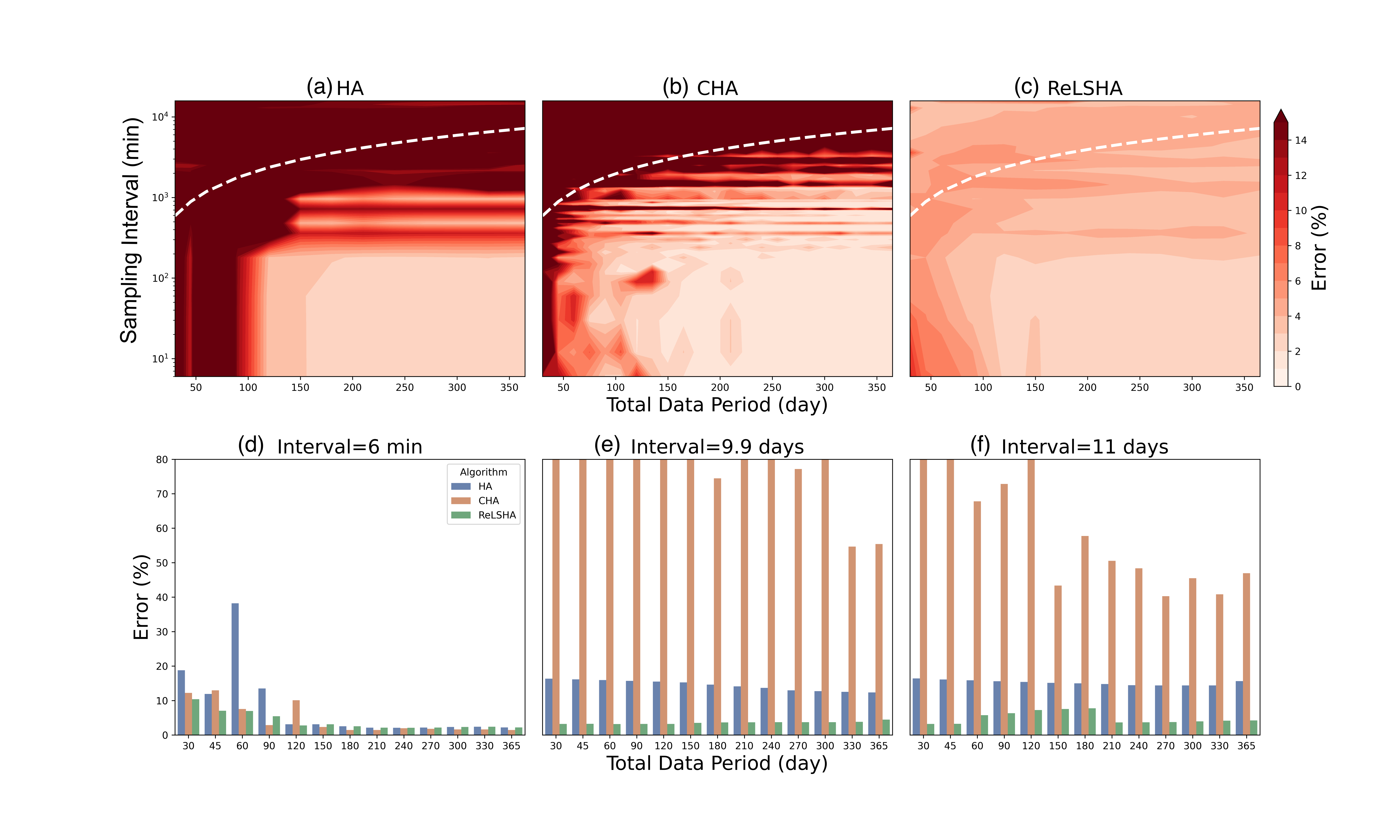}
    \caption{The comparison of the three harmonic analysis methods. (a-c) the RRMSE of HA, CHA, and ReLSHA methods (the white dashline marks the boundary between overdetermined and underdetermined regions); (d-f) the error for the three methods for comparison with (d) in the overdetermined region and (e) and (f) in the underdetermined region.}
    \label{fig:ha}
\end{figure}

The proposed ReLSHA is shown to outperform the traditional HA and CHA in Figure \ref{fig:ha}. Figure \ref{fig:ha}a-c show that the tidal amplitude RRMSE for ReLSH is generally much lower than the other two algorithms. Specifically, the range of the examination can be divided into two regions depending on the number of unknowns and measurements: ``overdetermined'' and ``underdetermined''. In the overdetermined region, the number of measurements is more than unknowns ($=37\times2=74$), while in the underdetermined region the number of unknowns is more than measurements. Traditional HA is required to only be applied in the ``overdetermined'' situation, otherwise the number of equations is less than the unknowns so that no unique solution can be obtained. We observed that low RRMSE is found for the overdetermined region where the sampling interval is short and the total data length is long. This result is expected, because the resampled data are close to meet the data requirements of sufficient total data lengths and short sampling intervals. In comparison, high RRMSE was found in the underdetermined region where the observation data points are less than the unknowns, i.e., there are infinite solutions to the HA analysis. Also, high RRMSE was found for the intervals of $\sim$12 hours, $\sim$24 hours, etc. in the results of HA and CHA, which is due to the problem that the sampling frequency is equal to the tidal constituent and the amplitude cannot be resolved. This issue was relatively trivial for ReLSHA – the RRMSE only showed peaks for three sampling frequencies. The RRMSE quickly increased for longer sampling intervals and shorter data periods for HA and CHA, whereas the error of ReLSHA remains relatively low in the range of the experiment. The highest error of Figure \ref{fig:ha}c is up to 10\% at the lower left corner, where the total data period is short and the sampling frequency is high. 

Similarly, the synthetic data experiment was performed in Galveston Bay, and the comparison among HA, CHA, and ReLSHA is shown in \ref{fig:galveston}b. HA had a low error when the sampling interval was short, but the error quickly grew when the interval was extended. CHA had a relative constant error over different sampling intervals, which reflected the difference between the reference sites of Morgans Point and Galveston Railroad Bridge and Eagle Point. Although ReLSHA had errors when the sampling interval was short compared to HA. It outperformed all the other methods when the sampling interval was extended to the level of days. The comparison further validated ReLSHA and showed that it could be applied to diurnal tides as well.

To reveal the error distribution and estimate the applicability of the algorithms to satellite data, a side-by-side comparison with three sampling intervals of 6 mins (the highest frequency), 9.9 days (revisit interval of Jason-3), and 11 days (the revisit interval of the new satellite, SWOT) was created in Figure \ref{fig:ha}d-f using the one-year data. At the sampling interval of 6 min, the error of all the methods declines quickly when the total data period increases, and CHA shows the best performance. However, at the intervals of 9.9 days and 11 days, the error of the CHA method is much higher than the others. HA maintains an error level of 15\%, while the proposed ReLSHA method could keep the error to $\sim$3\%. This comparison shows that in addition to achieving high accuracy, the ReLSHA method can even reach the capability to process the satellite altimetry data. In other words, ReLSHA with the input of prior tidal information of the region can break the constraints by the Nyquist-Shannon theorem and Rayleigh criterion to determine the tidal amplitudes and track the changes within an acceptable error tolerating the data quality of the satellite altimetry.

The second data experiment shows that ReLSHA is still the best algorithm to determine tidal ampltitudes (Figure \ref{fig:galveston}). Specifically, the result of the HA method had a rapid increase in RRMSE when the interval of the sampling increased. Since the two neighboring sites used in CHA are located in the same bay and have similar tidal characteristics to the targeted Eagle Point, the result of CHA had a constant RRMSE despite it being much lower than HA. ReLSHA had relatively higher but acceptable RRMSE when the sampling interval is short, but the advantage is shown in the range when sampling interval is long and available data points are scarce. Overall, ReLSHA showed the best performance.

\begin{figure}
    \centering
    \includegraphics[width=0.98\linewidth]{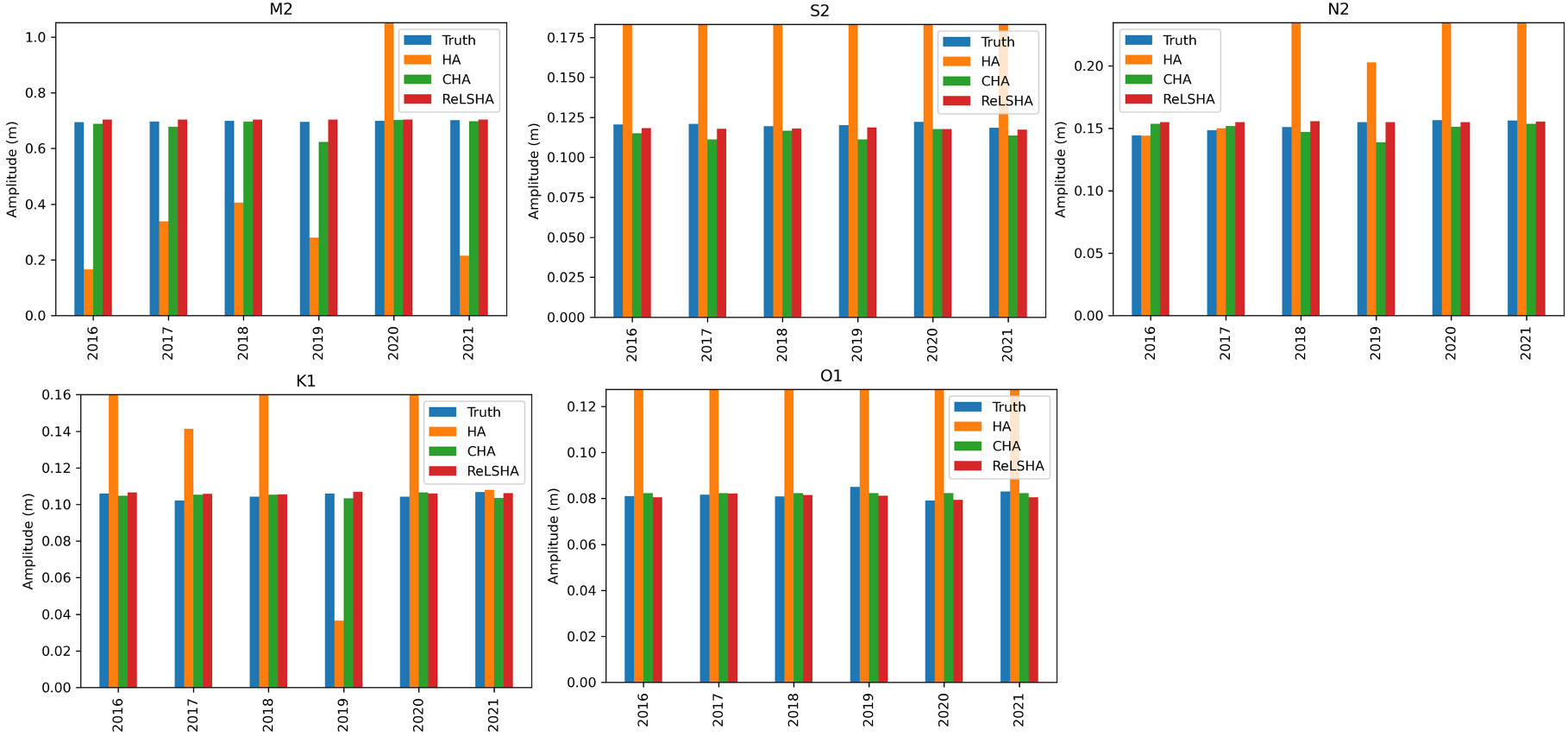}
    \caption{The comparison among HA, CHA, and ReLSHA based on the tidal amplitude measurements using Jason-3 altimetry data near the NOAA tidal gauge at Cape May.}
    \label{fig:trend}
\end{figure}

Encouraged by the high performance, we applied ReLSHA to the Jason-3 data from 2016 to 2021 and compared the result with the ground truth. The ground truth data is obtained using the traditional HA for the water level data of the NOAA tidal gauge at Cape May sampled at the 6-min interval for each year from 2016 to 2021. The analysis is shown in Figure \ref{fig:trend}, where ReLSHA is also compared with HA and CHA. For the five leading constituents, the ReLSHA-processed Jason-3 result is close to the ground truth. CHA has similar performance but has large errors for specific years and constituents, e.g., M2 in 2019. The similar performance of these two methods resulted from the selection of the reference site, BSL, a close tidal gauge to Cape May. The traditional HA method has much more significant errors compared to the other two methods. This demonstrates that ReLSHA has an excellent performance in processing real satellite data in spite of the noise, irregularity, and non-tidal signals in the data.

The critical component in ReLSHA is the selection of the reference tidal amplitudes. The present study focused on the tidal dynamics in Delaware Bay and Galveston, and the present result shows an exceptional performance even with a reference site far from the altimetry location. Further validation with other hydrodynamic conditions is required to test the generality of the method. In processing the real data such as from the Jason missions, the performance might be compromised due to the extremely long sampling interval and missing cycles. We would suggest using caution to apply ReLSHA with the best search of the reference information. As a general guide to improve the regularization, the reference site should be located with the most similar hydrodynamic characteristics -- sometimes the nearest site might not have similar hydrodynamics in a complicated tidal basin. At last, it is worth noting that validated numerical models could provide good reference information to serve as virtual reference sites. A future study, to test whether the ReLSHA result can be further improved using coastal hydrodynamics models, is being planned.

\section{Conclusion}
A new algorithm is proposed to accomplish a long-lasting, challenging task: how to use satellite altimetry data to estimate tidal amplitudes. The proposed algorithm of ReLSHA based on the Regularized Least-Square scheme provides a potential solution. This new method is designed to fit the tidal harmonics to the observation data with a penalization to the tidal amplitude difference between a reference and the harmonics. Synthetic data experiments using resampled NOAA tidal gauge data were conducted to validate the proposed algorithm, which shows much lower error than the traditional harmonic analysis and the best existing method, CHA. Since the testing sites are located strategically in different tidal basins with a long separation distance and different hydrodynamic conditions, the success of the synthetic data experiment showed that the tidal amplitudes along the Contiguous United States shoreline could potentially be determined with reasonable accuracy using the proposed algorithm if the satellite altimetry data are available. The application of ReLSH to the Jason-3 data further validated the method using real satellite altimetry data. The proposed method has great potential to increase the coverage and resolution of tidal amplitude observation worldwide on a yearly basis, especially in places where monitoring infrastructure is lacking such as developing countries. 

\acknowledgments
This project was partially supported by the U.S. Department of Transportation, Office of the Assistant Secretary for Research and Technology (OST-R), Grant No. 69A3551847102, issued to Rutgers, The State University of New Jersey and the Rutgers EOAS seeding grant. 

\section*{Open Research}
The tidal level data and Jason-3 data used for the validation in the study are available at figshare.com via https://doi.org/10.6084/m9.figshare.20076107.v1 with Attribution 4.0 International License (CC BY 4.0).


%
%
\bibliography{references.bib}
\nocite{wang_data_2022}

%
%
%
%
%

\end{document}